\documentclass[pdflatex,sn-mathphys]{sn-jnl}% Math and Physical Sciences Reference Style
\usepackage{array}
\usepackage{tabularx}
\usepackage{multirow}
\usepackage{array}
\usepackage{makecell}
\usepackage{subcaption}
\jyear{2022}%

\theoremstyle{thmstyleone}%
\theoremstyle{thmstyletwo}%

\theoremstyle{thmstylethree}%

\begin{document}

\title[Direct Observations of X-Rays Produced by Upward Positive Lightning]{Direct Observations of X-Rays Produced by Upward Positive Lightning}

%\title[First direct observations of X-rays associated with upward negative leader stepping]{First measurements of X-rays produced by positive cloud-to-ground lightning}


\author*[1]{\fnm{Toma} \sur{Oregel-Chaumont}}\email{toma.chaumont@epfl.ch}

\author[1]{\fnm{Antonio} \sur{\v{S}unjerga}}\email{antonio.sunjerga@epfl.ch}

\author[2]{\fnm{Pasan} \sur{Hettiarachchi}}\email{pasan.hettiarachchi@angstrom.uu.se}

\author[2]{\fnm{Vernon} \sur{Cooray}}\email{vernon.cooray@angstrom.uu.se}

\author[3]{\fnm{Marcos} \sur{Rubinstein}}\email{marcos.rubinstein@heig-vd.ch}

\author[1]{\fnm{Farhad} \sur{Rachidi}}\email{farhad.rachidi@epfl.ch}

\affil*[1]{\orgdiv{Electromagnetic Compatibility Laboratory}, \orgname{EPFL}, \orgaddress{\street{}, \city{Lausanne}, \postcode{1015}, \state{VD}, \country{Switzerland}}}

\affil[2]{\orgdiv{Department of Engineering Sciences}, \orgname{Uppsala University}, \orgaddress{\street{}, \city{Uppsala}, \postcode{751}, \state{}, \country{Sweden}}}

\affil[3]{\orgdiv{IICT}, \orgname{HEIG-VD}, \orgaddress{\street{}, \city{Yverdon-les-Bains}, \postcode{1401}, \state{VD}, \country{Switzerland}}}

%\affil[3]{\orgdiv{Department}, \orgname{Organization}, \orgaddress{\street{Street}, \city{City}, \postcode{610101}, \state{State}, \country{Country}}}


\abstract{X-rays have been observed in natural downward cloud-to-ground lightning for over twenty years and in rocket-triggered lightning for slightly less. 
In both cases, this energetic radiation has been detected during the stepped and dart leader phases of downward negative flashes. 
More recently, X-rays have also been reported during the dart leader phase of \textit{upward} negative flashes. 
In this study, we present the observations of four upward positive lightning flashes from the S\"antis Tower (2.5 km ASL) in Switzerland. 
These consist of the simultaneous records of electric current passing through the tower, and electric field strength and X-ray flux 20 meters from the tower base.
%Observations from a 2D interferometer system are also available for
One of the flashes was captured by a high-speed camera operating at 24,000 frames per second, stills from which are also presented. 
We detected X-rays during the initial phase of upward negative leader propagation, which can be associated with the leader-stepping process from electric field and current waveforms.
To the best of our knowledge, this is the first time that such measurements are reported in the literature. The obtained time-synchronised data confirm that the X-ray emissions detected are associated with the initial steps of the upward negative leader. The frequency and energy of X-ray pulses appear to decrease as a function of time, with pulses disappearing altogether within the first millisecond of the leader initiation. The X-ray pulse energy appears to increase with the maximum current-derivative and the electric field change of its associated leader step.
%In one of the four flashes, X-rays associated with the very first step of the upward negative leader have been observed. 
These observations contribute to improving the currently lackluster understanding of upward lightning, which is a primary source of damage to tall structures such as telecommunications towers and wind turbines, as well as airplanes during take-off and landing.}
%and the mechanisms involved in the initial breakdown.

%%================================%%
%% Sample for structured abstract %%
%%================================%%

% \abstract{\textbf{Purpose:} The abstract serves both as a general introduction to the topic and as a brief, non-technical summary of the main results and their implications.
% 
% \textbf{Methods:} The abstract must not include subheadings (unless expressly permitted in the journal's Instructions to Authors), equations or citations.
% 
% \textbf{Results:} As a guide the abstract should not exceed 200 words.
% 
% \textbf{Conclusion:} Most journals do not set a hard limit however authors are advised to check the author instructions for the journal they are submitting to.}




\maketitle

%-------------------------------% INTRO

\section{Introduction}\label{intro}

%The Introduction section, of referenced text \cite{bib1} expands on the background of the work (some overlap with the Abstract is acceptable). The introduction should not include subheadings.

%Springer Nature does not impose a strict layout as standard however authors are advised to check the individual requirements for the journal they are planning to submit to as there may be journal-level preferences. When preparing your text please also be aware that some stylistic choices are not supported in full text XML (publication version), including coloured font. These will not be replicated in the typeset article if it is accepted. 

The first unambiguous observation of X-ray generation from lightning flashes was made by Moore \textit{et al}. 2001 \cite{moore_energetic_2001}, who recorded X-ray bursts with energies in excess of 1 MeV during the stepped-leader phase of three natural downward negative lightning flashes. 
This confirmed the relativistic runaway electron avalanche (RREA) model proposed by Gurevich \textit{et al}. 1992 \citep{gurevich_runaway_1992}, which predicted the production of X-rays via the \textit{bremsstrahlung} interaction of electrons with air.
Since then, X-ray emissions have been measured in both natural and artificially-triggered cloud-to-ground (CG) lightning via a series of experiments conducted at Camp Blanding, Florida \citep{dwyer_energetic_2003, dwyer_x-ray_2005, saleh_properties_2009, mallick_study_2012}. 
Bursts of energetic radiation were detected during both the stepped-leader phase and dart leader--return stroke transition, with energies ranging from 100s of keV to 10s of MeV.
%X-rays appear in bursts on the order of a microsecond, with a typical total duration of about 1 ms or less, and carry up to MeVs of energy.

Measurements of X-ray emissions from natural upward lightning, however, were scanty until recently. 
Yoshida \textit{et al}. 2008 \cite{yoshida_high_2008} observed increased counts associated with seven lightning flashes on their plastic and NaI scintillators designed for detecting high-energy electron and photon bursts, though the 1-ms sampling interval of their detectors did not permit precise identification of the emitting phase. 
Out of the seven, they reported results on two, an upward negative flash and an upward positive flash.  
Montany\`a \textit{et al}. 2014 \cite{montanya_registration_2014} made measurements of X-ray emissions from several upward lightning flashes from the mountaintop Eagle Nest tower located at 2537 m above sea-level (ASL) in the Pyren\'ees.
They observed a 17 X-ray pulse burst (with an 806 keV maximum) during the stepped leader phase of a natural downward negative flash, but did not detect X-ray emissions during the 13 upward-initiated flashes reported.
Hettiarachchi \textit{et al}. 2018 \cite{hettiarachchi_x-ray_2018} were the first to directly measure X-ray emissions from upward-initiated lightning flashes: though either rare or very weak, they detected X-rays with energies up to 700 keV occurring both in bursts and as single events during the dart/dart-stepped leader phase of 3 natural upward negative flashes at Gaisberg Tower in Austria.

Herein we report, to the best of our knowledge, the first association of X-rays with the stepping of the upward negative leader in upward positive lightning flashes, as measured by the comprehensive S\"antis lightning measurement system. 
The data consist of simultaneous records of lightning current and its derivative, near electric field (20 m), and high-speed camera (HSC) images.
A summary of  the data types analysed in this study and the aforementioned studies is presented in Table~\ref{tab:lmscomp}.

%The overall observed increase in the incidence of lightning as a result of climate change highlights the importance of better characterising this .

%\begin{itemize}
%    \item X-ray measurements allow for better characterisation of the phenomenon;
%\end{itemize}

%-------------------------------% TABLE 1

\begin{table}[h!]
\begin{center}
%\begin{minipage}{<preferred-table-width>}
\caption{Lightning X-ray measurement studies -- A comparison}\label{tab:lmscomp}%
\begin{tabular}{ c | c | c | c | c | c } %{0.94\textwidth}
\toprule
\textbf{Study} & E-Field & Current & \multirow[t]{2}{*}{\makecell{High-speed \\ camera}} & \multirow[t]{2}{*}{\makecell{Interfer \\ -ometer}} & \textbf{Scintillators} \\
\midrule
Moore \textit{et al}. 2001 & YES & YES & NO & NO & 1 \\ \hline % South Baldy
Dwyer \textit{et al}. 2003-5 & YES & YES & YES & NO & 12 \\ \hline % ICLRT
Yoshida \textit{et al}. 2008 & YES & YES & NO & YES & 2 \\ \hline % Hokuriku
Saleh \textit{et al}. 2009 & YES & YES & NO & NO & 45 \\ \hline % TERA
Mallick \textit{et al}. 2012 & YES & NO & NO & NO & 1 \\ \hline % Gainesville
Montany\`a \textit{et al}. 2014 & YES & NO & YES & NO & 1 \\ \hline % Eagle Nest
Hettiarachchi \textit{et al}. 2018 & YES & YES & NO & NO & 2  \\ \hline % Gaisberg
This study & YES & YES & YES & NO\footnote{Interferometric (IFM) data is available but not analysed in this study.} & 2 \\
\botrule
\end{tabular}
%\end{minipage}
\end{center}
\end{table}

%-------------------------------% METHODS

\section{Methods}\label{sec:me}

%Topical subheadings are allowed. Authors must ensure that their Methods section includes adequate experimental and characterization data necessary for others in the field to reproduce their work. Authors are encouraged to include RIIDs where appropriate. 

\begin{figure}
    \centering
    \includegraphics[width=0.97\textwidth]{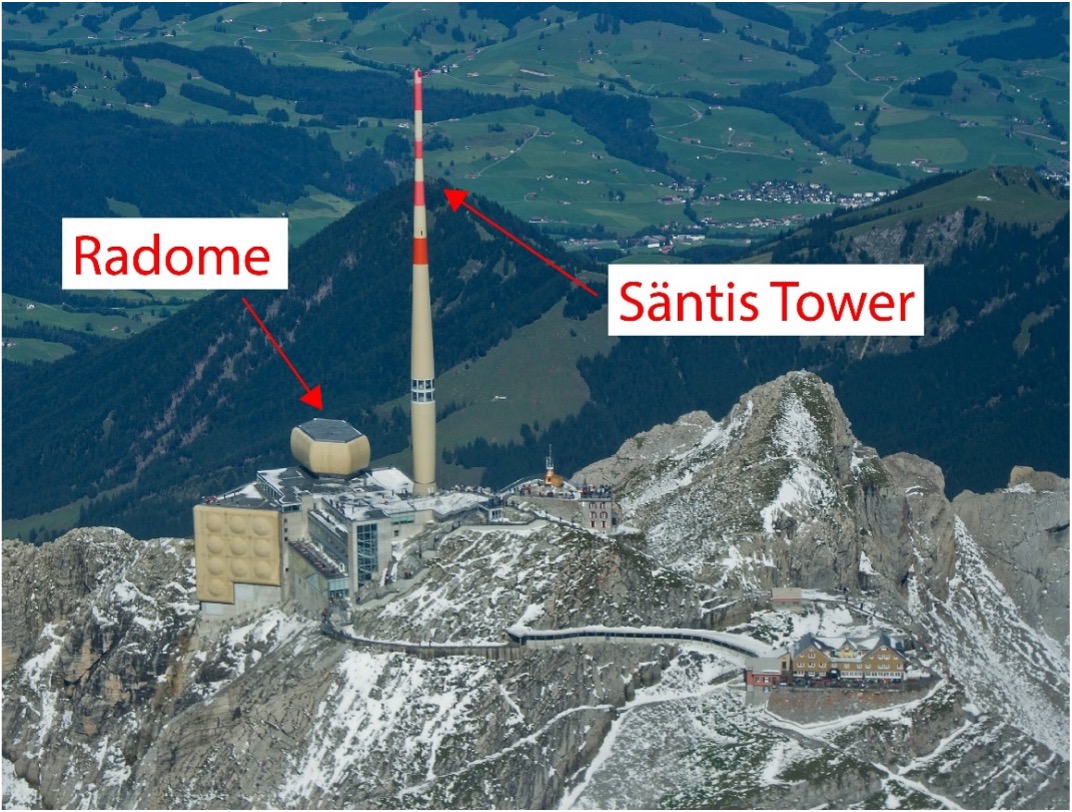}
    \caption{Photo of the S\"antis peak, with arrows indicating the Radome, which houses the electric field probe and scintillators, and the Tower, where the current and current-derivative sensors are located.}
    \label{fig:SantisPic}
\end{figure}

The Mt. S\"antis Lightning Research Facility, shown in Figure~\ref{fig:SantisPic}, is situated at 2502 m ASL in the Appenzell Alps of northeastern Switzerland, and experiences $>$100 direct lightning strikes per year to its 124 meter-tall tower, which is equipped with a Rogowski coil -- $\dot{B}$ sensor pair at two different heights (24 and 82 meters above ground level), for measurement of the current and current derivative, respectively.\footnote{All four sensors operate with a sampling rate of 50 MHz.}
The nearby Radome houses a near E-field sensor and two X-ray detectors (described below), which have a common sampling rate of 20 MHz.
Our M\'elop\'ee fast E-field probe has a frequency range of 1 kHz to 150 MHz and is described in more detail in \v{S}unjerga \textit{et al}. 2021 \cite{sunjerga_santis_2021}. 
Five kilometers away, atop Mt. Kr\"onberg (1663 ASL), is a high-speed camera (HSC) operating at 24,000 fps, with an exposure time of 41 $\mu$s.
Electric field measurements are also taken 15 km away by a flat-plate antenna with line-of-site in Herisau, Switzerland, though this data is not presented herein.
Additionally, during the Summer of 2021, when the flashes discussed below occurred, a University of New Mexico interferometer (IFM) was installed in Schw\"agalp, at the base of Mt. S\"antis. These interferometric results will be the subject of a separate paper.

One of our two NaI X-ray detectors belong to Uppsala University and the other to the University of California -- Santa Cruz (UCSC); the former records waveforms and is triggered by the tower current, whereas the latter records the peak energies of events and is working continuously.
The Uppsala scintillator, whose data are presented in this report, has a measuring range of $\sim$10 keV to 2 MeV, a temporal resolution of $\sim$1/4 $\mu$s, and is of the same design as that used by Hettiarachchi \textit{et al}.; refer to their 2018 paper \cite{hettiarachchi_x-ray_2018} for a detailed description.
%% Shielding of varying thickness to distinguish between single photons a multiple with window
This X-ray detector is connected to the same digitiser as our radome E-field probe and therefore by default synchronised; synchronisation of these two with the tower current and current derivative signals (which are themselves synchronised in the same manner) is done by aligning the time of the first E-field ``step'' with the time of the $\dot{B}$ extremum associated with the first current pulse, as these tend to be the sharpest.
The HSC and ``far'' E-field data are synchronised with the rest by GPS time-stamp if the antennae are functional at the time of the flash. If not, manual synchronisation can be carried out via waveform matching. More detailed information on the S\"antis measurement system can be found in  \cite{rachidi_santis_2022}.
%Mention any differences in band-filtering that could allow for differentiation between single MeV photon ($\gamma$-ray) and multiple $\sim$100 keV photons (X-rays).

All computational data analysis and presentation were carried out using the Python programming language, with the NumPy, SciPy and Matplotlib libraries in particular.

%-------------------------------% RESULTS

\section{Results}\label{sec:res}

%-------------------------------% TABLE 2

\begin{table}[h!]
\begin{center}
%\begin{minipage}{<preferred-table-width>}
\caption{Positive Lightning Flashes Analysed -- Data Summary}\label{tab:flashes}
\begin{tabular}{ c | m{5em} | c | c | c | c | c } %{tabular}{\textwidth}
\toprule
\textbf{Flash} & \textbf{Date \newline UTC} & \multirow[t]{2}{*}{\makecell{Prior \\ Activity}} & \multirow[t]{2}{*}{\makecell{Current \\ (\& derivative)}} & \multirow[t]{2}{*}{\makecell{High-speed \\ camera}} & \multirow[t]{2}{*}{\makecell{Interfer \\ -ometer}} & \multirow[t]{2}{*}{\makecell{E-field\footnote{Only near E-field waveforms are presented and analysed herein.} \\ (20-m \& 15-km)}} \\ % L#
\midrule
%UPa & 2021-06-28 \newline 23:08:40 &  & + & NO & NO & YES \\ \hline
UP0 & 2021-06-28 \newline 23:26:53 & YES & YES & NO & NO & YES \\ \hline % UPb
UP1 & 2021-07-24 \newline 16:06:07 & NO & YES & NO & YES & YES \\ \hline % L1
UP2 & 2021-07-24 \newline 16:24:03 & NO & YES & YES & NO & YES \\ \hline % L2
%UN1 & 2021-07-30 \newline 15:17:09 &  & -- & NO & YES & YES \\ \hline
%UN2 & 2021-07-30 \newline 15:30:51 &  & -- & NO & YES & NO \\ \hline
%UN3 & 2021-07-30 \newline 15:35:41 &  & -- & NO & YES & YES \\ \hline
%UN4 & 2021-07-30 \newline 15:38:10 & NO & -- & YES & YES & YES \\ \hline
UP3 & 2021-07-30 \newline 18:00:10 & YES & YES & NO & YES & YES \\ %\hline % L3
%UP4 & 2021-07-30 \newline 18:04:53 & YES & + & NO & YES & YES \\ %\hline % N06
%UN5 & 2021-08-16 \newline 02:34:10 &  & -- & NO & YES & YES \\ \hline
%UN6 & 2021-08-16 \newline 05:53:24 &  & -- & NO & YES & YES \\ \hline
%UN7 & 2021-08-16 \newline 15:06:45 &  & -- & NO & YES & NO \\ \hline
%UN8 & 2021-08-16 \newline 15:08:34 &  & -- & NO & YES & NO \\ %
\botrule
\end{tabular}
%\footnotemark[2] 
%\footnotetext[2]{}
%\end{minipage}
\end{center}
\end{table}

%-------------------------------%

We analysed 4 upward positive and 8 upward negative flashes with associated X-ray emissions that occurred during the Summer 2021 thunderstorm season.\footnote{A comprehensive analysis of all 12 flashes will be the subject of a separate paper.}
The data available for the 4 upward positive flashes (UPFs) presented here are summarised in Table~\ref{tab:flashes}.
In addition to tower current and electric field measurements, two (UP1 and UP3) were recorded by the interferometer (subject to a separate analysis), and one (UP2) was captured by the high-speed camera.
UP0 and UP3 also saw preceding lightning activity in the vicinity (i.e. intra-cloud flashes); studies have shown that this may impact the formation of leaders from the strike object \cite{sunjerga_initiation_2021}.
It should be noted that flashes UP1, UP2, and UP3 occurred during the Laser Lightning Rod project presented in Houard \textit{et al}. 2023 \cite{houard_laser-guided_2023} (therein called L1, L2, and L3, respectively), while the laser was on, whereas flash UP0 did not.

Here, we define the initial continuous current (ICC) at the start of the leader, i.e., the first significant deviation from zero of the electric field, current and current-derivative. 
For the latter two, we have chosen the convention of a negative current corresponding to a positive charge transfer from cloud to ground.
Each positive flash had between two and seven X-ray events associated with this ``stepping'' of the upward negative leader, also indicated by pulses in the current waveform.

%(on average approximately 3 per UN).
%Note that upward negative lightning is characterized by a higher number of current pulses as compared to downward lightning.

%-------------------------------% TABLE 3

\begin{table}[h!]
\begin{center}
\caption{Positive flash pulses with associated X-rays}
\begin{tabular}{ c | c | c | c | c | c | c | c }
\toprule
\textbf{Flash} & $t_{SL}$ [$\mu$s] & $I_p$ [kA] & $t_{mr}$ [$\mu$s] & \textbar$\frac{dI}{dt}$\textbar\, [$\frac{\mathrm{kA}}{\mu\mathrm{s}}$] & $\Delta E$ [$\frac{\mathrm{V}}{\mathrm{m}}$] & $t_{Er}$ [$\mu$s] & XRE [keV] \\
\midrule %\hline %
%\multirow{1}{*}{UPa} & 778.2 & 1.17$\pm$0.05 & 0.87$^{+0.30}_{-0.19}$ & 1.3$\pm$0.3 & 215$\pm$55 & 2.95 & 176$\pm$1 \\ \hline
\multirow{1}{*}{UP0} & 150.9 & 0.81$\pm$0.05 & 0.13$^{+0.02}_{-0.01}$ & 6.2$\pm$0.4 & 545$\pm$55 & 0.15 & 63.4$\pm$0.9 \\ \hline
\multirow{7}{*}{UP1} & 0.0 & 1.66$\pm$0.05 & 0.12$^{+0.01}_{-0.01}$ & 14.2$\pm$0.3 & 1025$\pm$40 & 0.20 & 259.3$\pm$0.7 \\ 
                     & 89.5 & 2.01 & 0.24$^{+0.02}_{-0.02}$ & 8.2 & 885 & 0.30 & 85.1 \\ 
                     & 334.6 & 1.91 & 0.88$^{+0.18}_{-0.14}$ & 2.2 & 635 & 9.50 & 87.0 \\ 
                     & 345.7 & 2.89 & 0.79$^{+0.09}_{-0.08}$ & 3.7 & 635 & 9.50 & 123.8 \\ 
                     & 465.6 & 2.20 & 1.53$^{+0.50}_{-0.32}$ & 1.4 & 330 & 10.55 & 31.6 \\  
                     & 554.1 & 2.80 & 1.36$^{+0.29}_{-0.21}$ & 2.1 & 590 & 17.10 & 79.8 \\ 
                     & 774.0 & 2.66 & 1.78$^{+0.55}_{-0.35}$ & 1.5 & 350 & 19.50 & 31.1 \\ \hline
\multirow{2}{*}{UP2} & 46.3 & 1.14$\pm$0.05 & 0.13$^{+0.01}_{-0.01}$ & 8.7$\pm$0.3 & 1450$\pm$40 & 0.20 & 54.6$\pm$0.7 \\ 
                     & 267.5 & 0.41 & 0.31$^{+0.14}_{-0.09}$ & 1.3 & 280 & 11.05 & 44.1 \\ \hline
\multirow{4}{*}{UP3} & 118.0 & 2.44$\pm$0.05 & 0.47$^{+0.04}_{-0.04}$ & 5.2$\pm$0.3 & 720$\pm$40 & 0.25 & 26.5$\pm$0.8 \\
                     & 355.0 & 2.71 & 1.28$^{+0.26}_{-0.19}$ & 2.1 & 475 & 2.35 & 50.4 \\
                     & 483.0 & 1.82 & 1.08$^{+0.30}_{-0.20}$ & 1.7 & 175 & 6.70 & 53.0 \\
                     & 785.0 & 5.20 & 2.52$^{+0.51}_{-0.37}$ & 2.1 & 565 & 21.10 & 42.8 \\ \hline
$\mu_a \pm \sigma_a$ & -- & 2.19$\pm$1.11 & 0.90$\pm$0.71 & 4.3$\pm$3.7 & 620$\pm$320 & 7.75$\pm$7.30 & 73.8$\pm$57.6 \\ \hline
$\mu_g  \sigma_g^{\pm 1}$ & -- & $1.88^{+1.55}_{-0.85}$ & $0.59^{+1.05}_{-0.38}$ &  $3.2^{+3.6}_{-1.7}$ & $540^{+375}_{-220}$ & $2.60^{+14.90}_{-2.20}$ & $60.4^{+48.0}_{-26.7}$ \\ %\hline
\botrule
\end{tabular}
\label{tab:pfxr}
\end{center}
\end{table}

%-------------------------------%

Table~\ref{tab:pfxr} presents the measured data for each pulse with correlated X-rays. 
The time $t_{SL}$ (``stepped leader'') is measured from the onset of the ICC. 
$I_p$ represents the absolute value peak current of a given pulse, and \textbar$
dI/dt$\textbar$_\mathrm{max} \,$ its maximum current derivative (slope). 
Together they make the minimum current rise-time, defined by Giri \textit{et al}. 2009 \cite{giri_relationship_2009} as:
\begin{equation}
    t_{mr} = \frac{I_p}{\|\frac{dI}{dt}\|_\mathrm{max}}
\end{equation}
which has a temporal accuracy of 20 ns.
The change in the electric field is given by $\Delta E$ and its 80\% rise-time by $t_{Er}$ (with a temporal resolution of 50 ns). 
Finally, XRE is the associated X-ray energy.
The first row of each flash provides the error associated with sensor noise at that time, with the exception of the calculated $t_{mr}$, whose errors vary with measurement.
The last two rows in the table provide the arithmetic and geometric means and standard deviations of each data set above. One can already see that all parameters except $I_p$ exhibit at least some degree of temporal variation, as will be confirmed later in Section~\ref{sec:dis}.

%-------------------------------% FIGURE 2 (UP1)

\begin{figure}[h!]
\centering
    \begin{subfigure}[t]{0.49\textwidth}
        \centering
        \includegraphics[width=\textwidth]{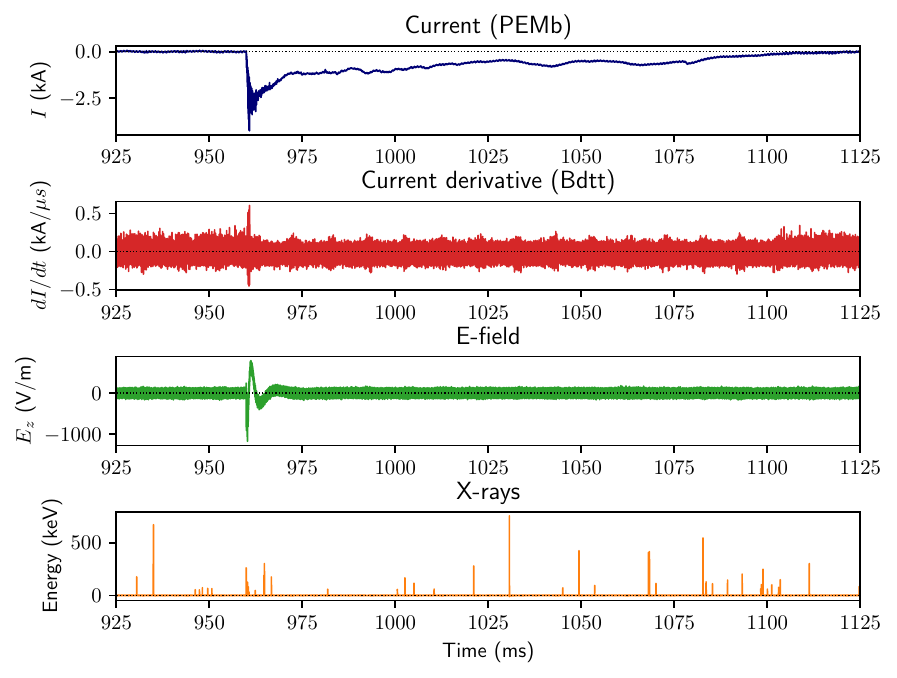}
        \subcaption{The entire duration of the flash. A 100 kHz low-pass filter has been applied to the current and $dI/dt$ waveforms to remove intermittent noise.}\label{fig:up1w}
    \end{subfigure}
    \hfill
    \begin{subfigure}[t]{0.49\textwidth}
        \centering
        \includegraphics[width=\textwidth]{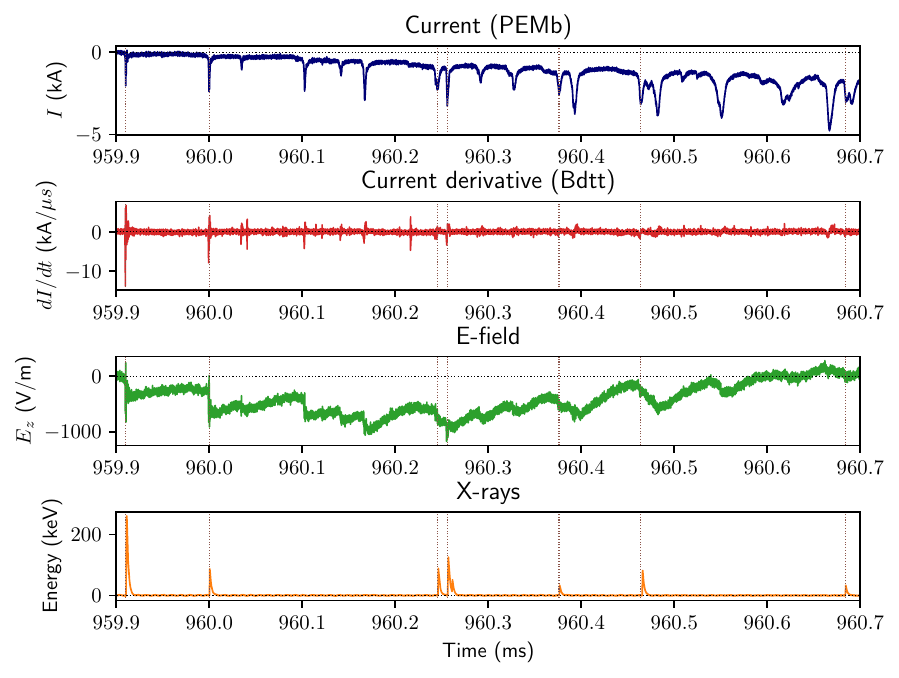}
        \subcaption{Zoom on the X-ray events during the upward stepping negative leader phase. The brown vertical dotted lines indicate the event times. See Table~\ref{tab:pfxr} for pulse data.}\label{fig:up1sl}
    \end{subfigure}
\caption{Waveforms of UP1, a Type 2 upward positive flash that occurred on July 24, 2021 at 16:06:07 UTC. ``PEMb'' and ``Bdtt'' specify the bottom Rogowski coil and top $\dot{B}$ sensor, respectively. $E_z$ is the measured vertical component of the electric field at 20 m. The time is from the beginning of the recording ($\sim$1 second before the current peak).}\label{fig:up1}
\end{figure}

%-------------------------------%

The current, $dI/dt$, near E-field, and X-ray waveforms of UP1 are shown in Figure~\ref{fig:up1}. 
It can be seen from Figure~\ref{fig:up1w} that this was a Type 2 upward positive flash, as defined by Romero \textit{et al}. 2013 \cite{romero_positive_2013}; i.e., it lacks a return stroke--like main pulse following the stepped-leader phase. 
Figure~\ref{fig:up1sl} presents an expanded view of the initiation of the upward leader and its stepping. 
It clearly shows how steps in the electric field are  associated with ICC pulses; 1/3 of which were associated with X-ray emissions. 
The 7 X-ray pulses have a median temporal separation on the order of 100 $\mu$s, and median energy on the order of 80 keV. 
Note, however, the decrease in pulse peak energy as time goes on.

%-------------------------------% FIGURE 3 (UP2 whole)

\begin{figure}[h!]
\centering
    \begin{subfigure}{0.56\textwidth}
        \centering
        \includegraphics[width=\textwidth]{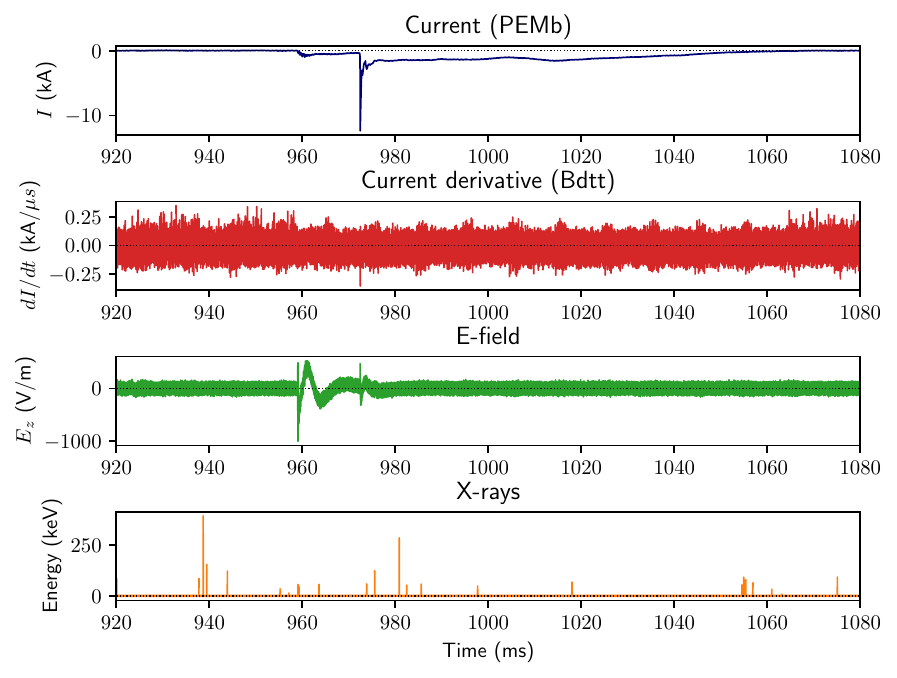}
    \end{subfigure}
    \hfill
    \begin{subfigure}{0.42\textwidth}
        \centering
        \includegraphics[width=\textwidth]{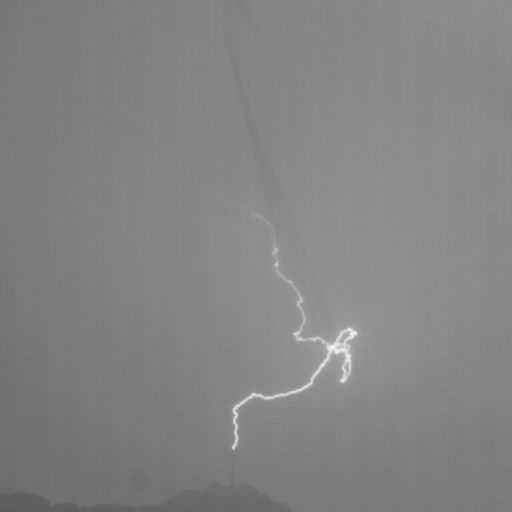}
    \end{subfigure}
\caption{Waveforms and integrated HSC frames of UP2, a Type 1 upward positive flash that occurred on July 24, 2021 at 16:24:03 UTC. A 100 kHz low-pass filter has been applied to the current and $dI/dt$ waveforms to remove intermittent noise. See Figure~\ref{fig:up2sl} for a zoom-in view on the X-ray events.}\label{fig:up2w}
\end{figure}

%-------------------------------%

\begin{figure}[h!]
\centering
    \begin{subfigure}{0.49\textwidth}
        \centering
        \includegraphics[width=\textwidth]{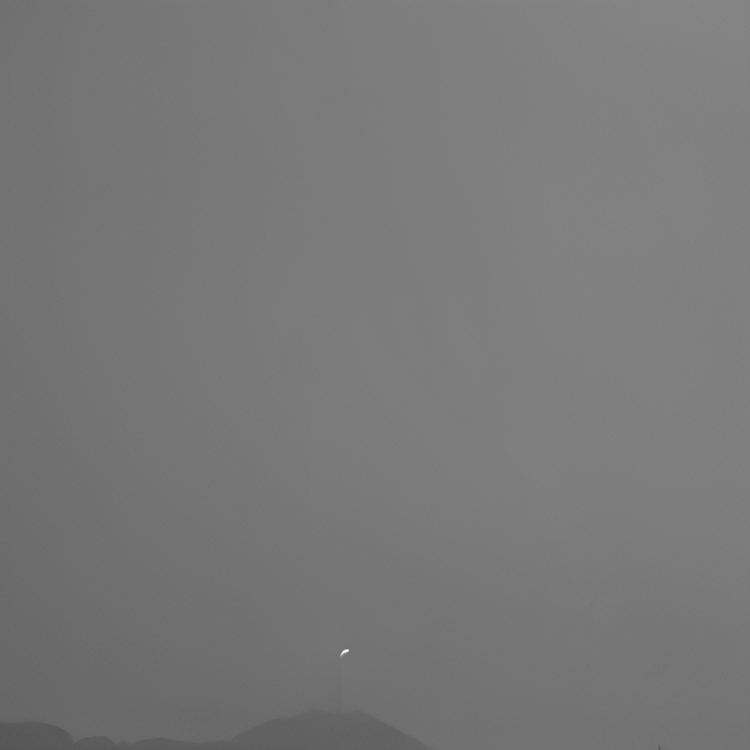}
        \subcaption*{HSC Frame 655}
    \end{subfigure}
    \hfill
    \begin{subfigure}{0.49\textwidth}
        \centering
        \includegraphics[width=\textwidth]{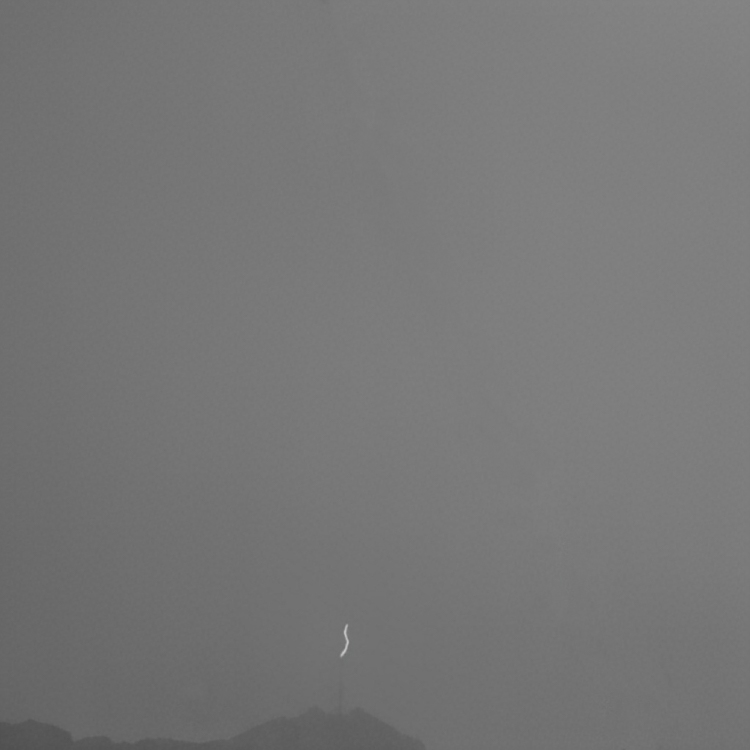}
        \subcaption*{HSC Frame 661}
    \end{subfigure}
    \begin{subfigure}{0.98\textwidth}
        \centering
        \includegraphics[width=\textwidth]{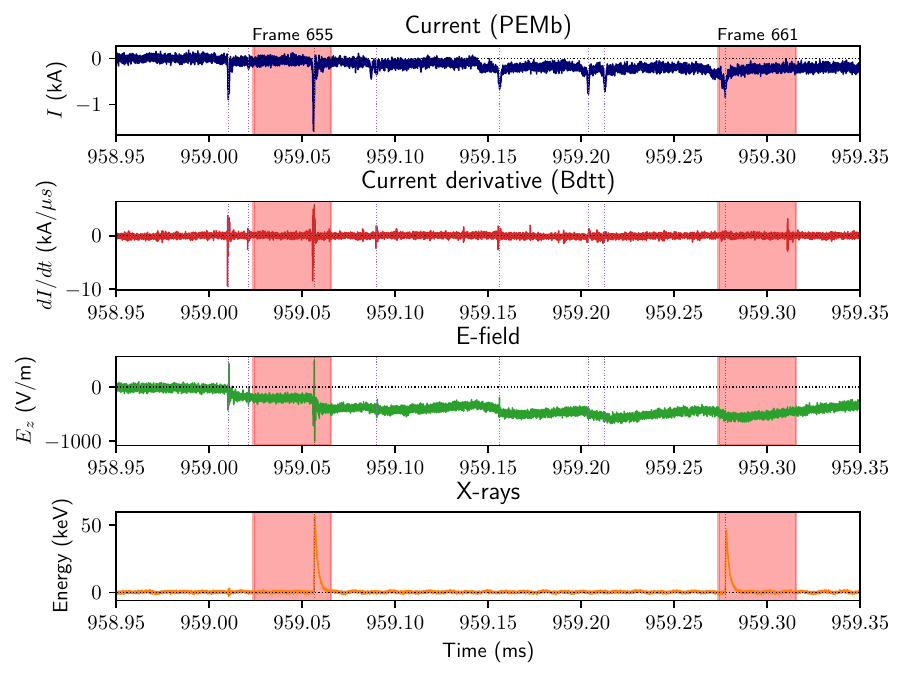}
    \end{subfigure}
\caption{X-ray events during the upward stepping negative leader phase of UP2. HSC frames containing the two X-ray pulses observed are shown above, and their approximate temporal width ($\sim$42 $\mu$s) is highlighted by the red-shaded regions in the waveforms below. E-field steps without associated X-rays are also indicated by the violet vertical dotted lines. See Table~\ref{tab:pfxr} for pulse data and Figure~\ref{fig:up2w} for a zoomed-out view of the waveforms.}\label{fig:up2sl}
\end{figure}

The whole flash waveforms and integrated HSC frames of UP2 are shown in Figure~\ref{fig:up2w}. 
This was clearly a Type 1 upward positive flash \cite{romero_positive_2013}, with a very obvious return stroke--like pulse after the upward-stepping leader. 
In the right-hand panel of the figure, one can make out the S\"antis tower, from which the flash initiated, at the base of the rather tortuous plasma channel.\footnote{The faint black streak running diagonally across the integrated stills are raindrops streaming down the camera's protective window pane.} 
The plots at the bottom of Figure~\ref{fig:up2sl} provide a zoom on the beginning of the ICC, when the two X-ray pulses occurred and the top pictures are HSC stills containing these pulses, which occurred 221 $\mu$s apart with an average energy of 49 keV.
Once again, these are clearly associated with the leader-stepping process, though unlike the Type 2 UPFs, only 1/10 of the leader steps had accompanying X-rays detected.

%-------------------------------% FIGURE 3 (UP2 events)

%-------------------------------%

UP3 was a Type 2 flash like UP1; its 4 X-ray pulses (about 1/6 of all ICC pulses) had a mean temporal separation of $\sim$220 $\mu$s and an average energy of 43 keV.
UP0 was likely a Type 1 flash like UP2; its singular X-ray pulse ($<$5\% of all ICC pulses) had an energy of 63 keV.
These flashes' waveforms are not depicted here for the sake of conciseness, though all their pulse data have been included in the following analysis.
See Appendix Figures~\ref{fig:UP0} \& \ref{fig:UP3} for plots.

\section{Discussion}\label{sec:dis}

%Discussions should be brief and focused. In some disciplines use of Discussion or `Conclusion' is interchangeable. It is not mandatory to use both. Some journals prefer a section `Results and Discussion' followed by a section `Conclusion'. Please refer to Journal-level guidance for any specific requirements. 

\begin{figure}[h!]
\centering
    \begin{subfigure}[t]{0.49\textwidth}
        \centering
        \includegraphics[width=\textwidth]{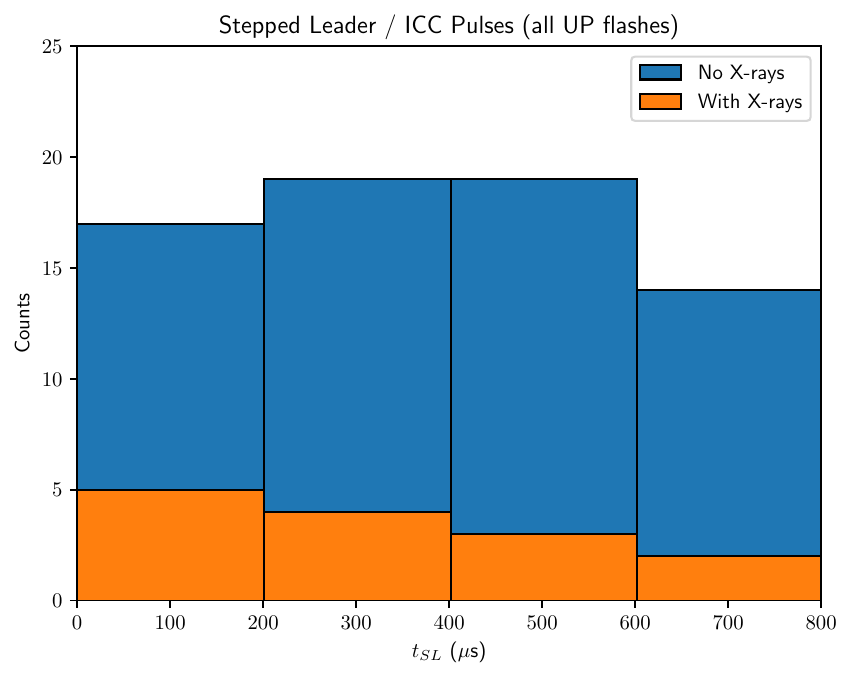}
        \subcaption{Histogram of ICC pulse counts, both with and without accompanying X-rays, as a function of time.}\label{fig:xrvth}
    \end{subfigure}
    \hfill
    \begin{subfigure}[t]{0.49\textwidth}
        \centering
        \includegraphics[width=\textwidth]{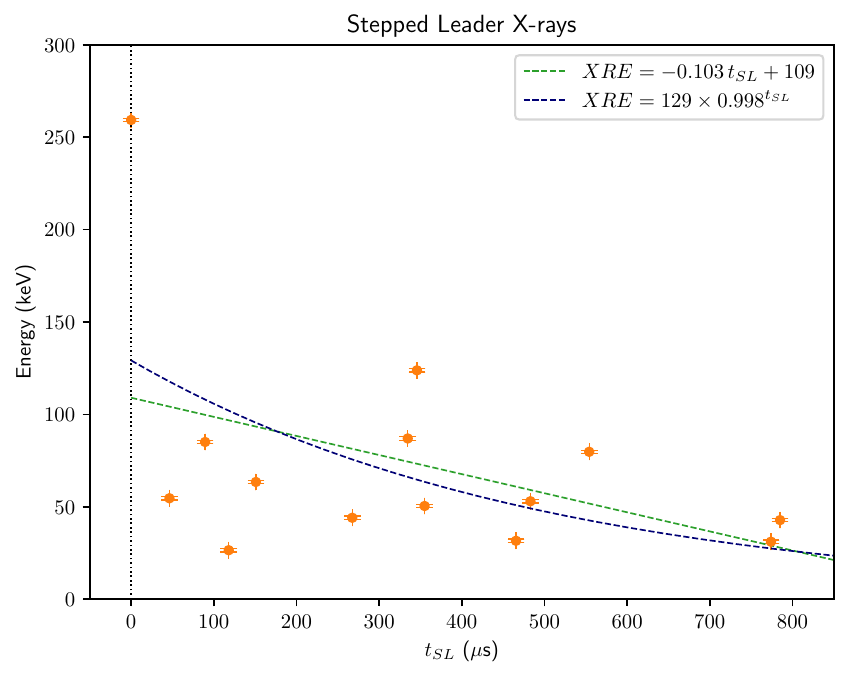}
        \subcaption{Scatter plot of X-ray energy as a function of time. The affine (green) and exponential (blue) fit lines have r$^2$ values of 0.19 and 0.24, respectively.}\label{fig:xrevt}
    \end{subfigure}
\caption{Plots depicting the temporal dependence of the X-ray counts and pulse energy for flashes UP0, UP1, UP2 and UP3. Time $t_{SL} = 0$ is set to the start of the stepped leader / ICC. Data taken from Table~\ref{tab:pfxr}.}\label{fig:xrvt}
\end{figure}

Figure~\ref{fig:xrvt} shows how both X-ray count and energy decrease as a functions of time from the onset of the stepped leader, $t_{SL}$. 
Note how, in comparison with the X-ray pulse to non-X-ray pulse ratios presented in Section~\ref{sec:res}, ICC pulses with measured accompanying X-rays compose a steadily decreasing percentage of all measured pulses, starting at $\sim$29\% during the first 200 $\mu$s, and dropping to 0\% after 800 $\mu$s.
Although one could argue that this count decrease observed in Figure~\ref{fig:xrvth} is simply due to the decrease in photon flux at the sensor location as the leader tip (where the X-rays are presumed to be emitted \cite{rakov_lightning_2003, moss_monte_2006}) moves away, the same argument cannot be made for observed \textit{energy} decrease in Figure~\ref{fig:xrevt}, as the waveforms are indicative of single events, rather than photon-burst energy pile-ups (compare with Figure 4 of Saleh \textit{et al}. 2009 \cite{saleh_properties_2009}).
The best-fit lines in this plot imply the existence of a finite time ($>$1 ms) after which X-rays would no longer be generated and/or detected.

%-------------------------------% FIGURE 5

%-------------------------------%

\begin{figure}[h!]
\centering
    \begin{subfigure}[t]{0.49\textwidth}
        \centering
        \includegraphics[width=\textwidth]{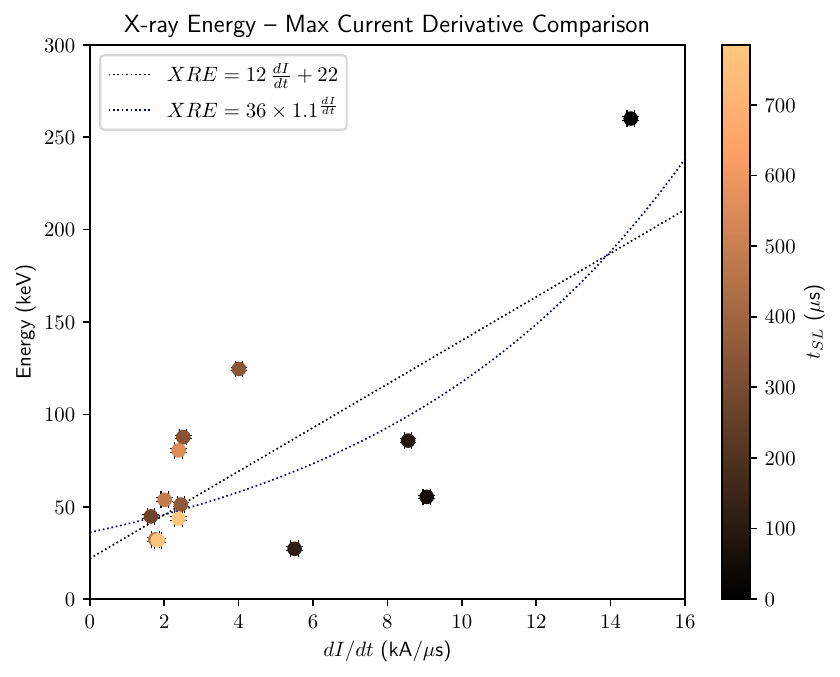}
        \subcaption{}\label{fig:xredidt}
    \end{subfigure}
    \hfill
    \begin{subfigure}[t]{0.49\textwidth}
        \centering
        \includegraphics[width=\textwidth]{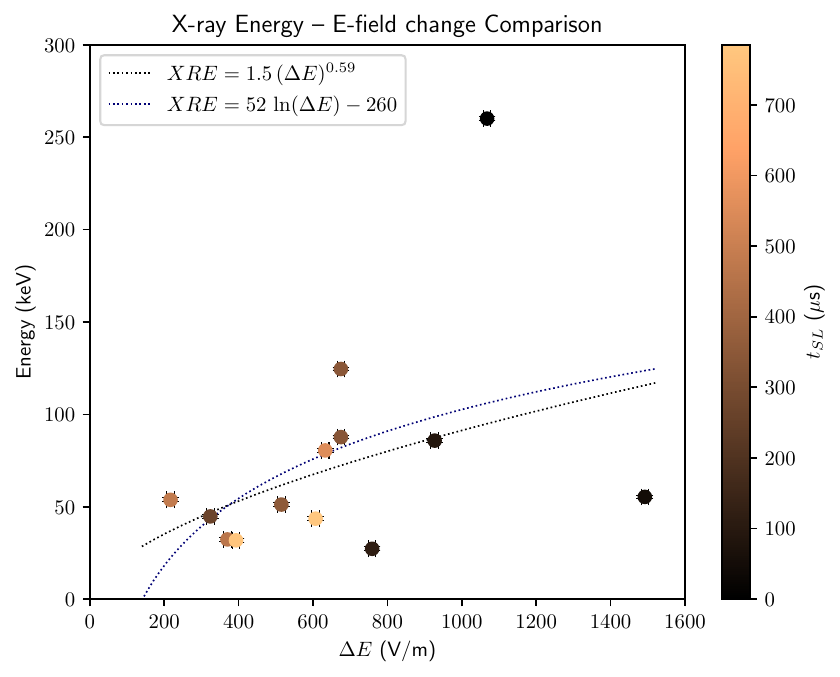}
        \subcaption{}\label{fig:xrede}
    \end{subfigure}
\caption{Color-mapped scatter plots depicting the parametric dependence of the X-ray energy for flashes UP1, UP2 and UP3, similar to Figure 15 of Mallick \textit{et al}. 2012 \cite{mallick_study_2012}. Time $t_{SL} = 0$ is set to the start of the stepped leader / ICC. Data taken from Table~\ref{tab:pfxr}.}\label{fig:xrevpar}
\end{figure}

Figure~\ref{fig:xredidt} shows the scatter plot of X-ray energy versus maximum current derivative, \textbar$\frac{dI}{dt}$\textbar$_\mathrm{max}$. 
It is clear from the color map that the latter also decreases as a function of time $t_{SL}$ (alternatively, the pulse rise-time $t_{mr}$ increases). 
The best fit lines show that the X-ray energy increases with \textbar$\frac{dI}{dt}$\textbar$_\mathrm{max}$, either linearly or exponentially. 
%Note that the y-intercept at $\sim$30 keV (with either fit) is of the same order as our scintillator's minimum threshold energy, and the background X-ray flux ($\sim$20 keV); in other words, it is not surprising that X-rays should be measured even in the absence of current pulses.

%-------------------------------% FIGURE 6

%-------------------------------%

Figure~\ref{fig:xrede} shows the scatter plot of X-ray energy versus electric field change, $\Delta E$. 
It is clear from the color map that $\Delta E$ decreases as a function of time, as does the X-ray energy, albeit to a lesser extent (see Figure~\ref{fig:xrevt}). 
The former can be at least partly explained by the fact that our sensor measures only the vertical component of the electric field, and the leader tip is moving away from the tower. 
The best fit lines show how the X-ray energy increases with $\Delta E$, possibly logarithmically or as a power-law. 
Note that, should a logarithmic relation be confirmed by further investigation, the x-intercept at $\Delta E \approx 140$ V/m implies the existence of a minimum E-field change needed for X-rays to be produced.

As such, it has been suggested (e.g., \cite{dwyer_implications_2004, cooray_electric_2009}) that the so-called ``cold runaway electron mechanism'', as opposed to the RREA model, is active in X-ray emissions associated with lightning leaders.
For the cold runaway mechanism to be active, it is necessary for the background electric field at atmospheric pressure to exceed about 20 MV/m \cite{moss_monte_2006, cooray_electric_2009}. 
If the electric field increases slowly in atmospheric air, as its value reaches around 3 MV/m the normal electrical breakdown process takes over and the resulting increase in conductivity of the discharge channel limits further increase of the electric field strength. 
The electric field will therefore be clamped to a value equal to or below this breakdown threshold.\footnote{If the pressure is below atmospheric, these mechanisms remain the same except that the threshold fields are scaled down linearly with pressure.} 
Since a certain amount of time is needed for the completion of standard breakdown, in order to achieve the cold runaway mechanism the electric field has to increase very rapidly in a given region of space so that there isn't sufficient time for the standard breakdown mechanism to take over and clamp the electric field at $\sim$3 MV/m. 
Thus, only very fast discharge processes (sub-microsecond scale) can generate the strong electric fields needed to push electrons into the cold runaway regime quickly enough \cite{cooray_electric_2009}. 
This is in agreement with the observation that X-ray emissions occur during discharge processes with rapidly changing currents, such as those seen in this study.

\section{Conclusion}\label{sec:con}

%Conclusions may be used to restate your hypothesis or research question, restate your major findings, explain the relevance and the added value of your work, highlight any limitations of your study, describe future directions for research and recommendations. 

%In some disciplines use of Discussion or 'Conclusion' is interchangeable. It is not mandatory to use both. Please refer to Journal-level guidance for any specific requirements. 

%In contrast to previous studies of X-rays associated with upward negative lightning we observed relatively high incidence of X-ray bursts. These X-rays bursts are associated with stepping of downward leader during dart leader phase. This is in line with recent observations from downward negative lightning. Our study suggests that production mechanism during dart leader phase is similar in both upward and downward lighting.

Herein we reported, to the best of our knowledge, the first measurements of X-rays produced by positive lightning flashes, specifically during the stepping of the upward negative leader.
We presented the waveforms of the current, current derivative, electric field, and X-ray energy for the four flashes in question (two Type 2 upward positive flashes), as well as high-speed camera stills for one of them (a Type 1 upward positive flash).
These time-synchronised data served to confirm that the X-ray emissions detected are associated with the initial steps of the upward negative leader.
Further analysis of the parameters at play revealed three additional points of interest:
\begin{itemize}
    \item The frequency and energy of X-ray pulses appear to decrease as a functions of time, with pulses disappearing altogether within the first millisecond of leader initiation;
    \item the Type 1 upward positive flashes exhibited the lowest percentages of pulses with accompanying X-rays, which also ended sooner;
    \item X-ray pulse energy appears to increase with the maximum current-derivative and the electric field change of its associated leader step. This supports the cold runaway electron model as the active mechanism for lightning leader X-ray production.
\end{itemize}

\noindent These observations contribute to improving our understanding of upward lightning, and will soon be followed by a more comprehensive review, including X-ray--emitting upward \textit{negative} flashes observed at the S\"antis tower, and simultaneous interferometric data gathered during the summer 2021 experimental campaign.

\backmatter

\bmhead{Supplementary information}

None.
%If your article has accompanying supplementary file(s) please state so here. Please refer to Journal-level guidance for any specific requirements.

\bmhead{Acknowledgments}

%Acknowledgments are not compulsory. Where included they should be brief. Grant or contribution numbers may be acknowledged.
%Please refer to Journal-level guidance for any specific requirements.

This work was supported in part by the Swiss National Science Foundation (Project no. 200020\_204235) and the European Union's Horizon 2020 research and innovation program (grant agreement no. 737033-LLR).
The authors would like to thank Florent Aviolat for developing a data-visualisation software that expedited the identification of events.%, and Yann-Cédric Bebene for his parallel analysis of the upward negative flashes discussed herein.

\section*{Declarations}

%Some journals require declarations to be submitted in a standardised format. Please check the Instructions for Authors of the journal to which you are submitting to see if you need to complete this section. If yes, your manuscript must contain the following sections under the heading `Declarations':

%\begin{itemize}
%    \item Funding
%    \item Ethics approval 
%    \item Consent to participate
%    \item Consent for publication
%\end{itemize}

%\noindent
%If any of the sections are not relevant to your manuscript, please include the heading and write `Not applicable' for that section. 

\subsection*{Competing interests}
None.

\subsection*{Data availability}
All processed data analysed during this study are included in this published article (and its supplementary information files). Raw data sets generated during the current study are available from the corresponding author on reasonable request.

\subsection*{Code availability}
Upon request from corresponding author.

\subsection*{Authors' contributions}
T.O.C. wrote the manuscript and completed the data analysis and processing begun by A.S., who also gathered the data alongside M.R. and F.R., heads of the S\"antis Lightning Research Facility, during the summer 2021 experimental campaign. P.H. and V.C. designed and built the X-ray detector providing the data.

%\bigskip


%\bigskip\noindent
%Springer journals and proceedings: \url{https://www.springer.com/gp/editorial-policies}

%\bigskip\noindent
%Nature Portfolio journals: \url{https://www.nature.com/nature-research/editorial-policies}

%\bigskip\noindent
%\textit{Scientific Reports}: %\url{https://www.nature.com/srep/journal-policies/editorial-policies}

%\bigskip\noindent
%BMC journals: \url{https://www.biomedcentral.com/getpublished/editorial-policies}
%\end{flushleft}

\begin{appendices}

\section{Flashes UP0 \& UP3}\label{app:plots}

% UP1 
% $\overline{E} = 20 \pm 50$ V/m. $E(t_\mathrm{min}) = -1156$ V/m. $E(t_\mathrm{max}) = 805$ V/m.}

%UP2
%$\overline{E} = 12 \pm 45$ V/m. $E(t_\mathrm{min}) = -992$ V/m. $E(t_\mathrm{max}) = 538$ V/m.}

%UP3
%$\overline{E} = 10 \pm 80$ V/m. $E(t_\mathrm{min}) = -2510$ V/m. $E(t_\mathrm{max}) = 2125$ V/m.

%-------------------------------% UP0 plots

\begin{figure}[H]
    \centering
    \begin{subfigure}[t]{0.49\textwidth}
        \centering
        \includegraphics[width=\textwidth]{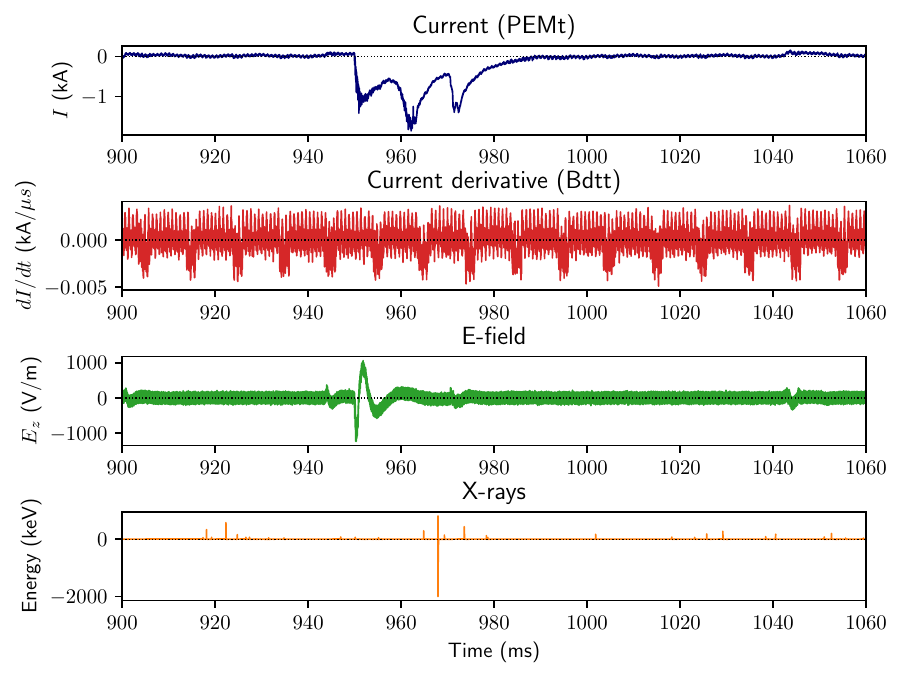}
        \subcaption{The entire duration of the flash. A 100 kHz low-pass filter has been applied to the current and $dI/dt$ waveforms to remove intermittent noise.}\label{fig:up0w}
    \end{subfigure}
    \hfill
    \begin{subfigure}[t]{0.49\textwidth}
        \centering
        \includegraphics[width=\textwidth]{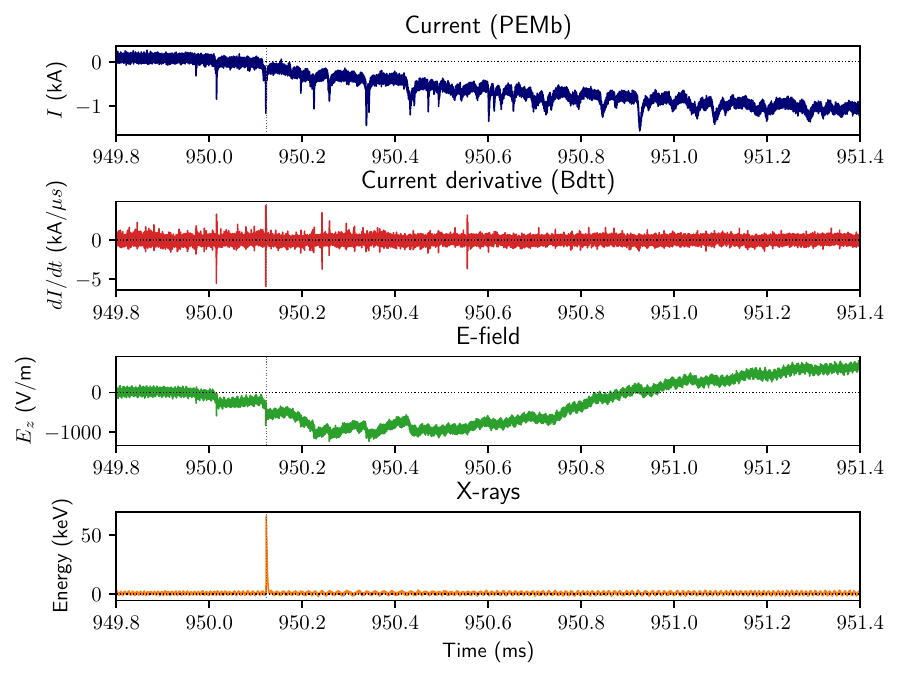}
        \subcaption{Zoom on the X-ray event during the upward stepping negative leader phase. The brown vertical dotted lines indicate the event times. See Table~\ref{tab:pfxr} for pulse data.}\label{fig:up0sl}
    \end{subfigure}
    \caption{Data associated with the Type 1 upward positive flash UP0, that occurred on June 28, 2021 at 23:26:29 UTC. ``PEMb'' and ``Bdtt'' specify the bottom Rogowski coil and top $\dot{B}$ sensor, respectively. $E_z$ is the measured vertical component of the electric field. The time is from the beginning of the recording ($\sim$1 second before the current peak).}
    \label{fig:UP0}
\end{figure}

%-------------------------------% UP3 plots

\begin{figure}[H]
    \centering
    \begin{subfigure}[t]{0.49\textwidth}
        \centering
        \includegraphics[width=\textwidth]{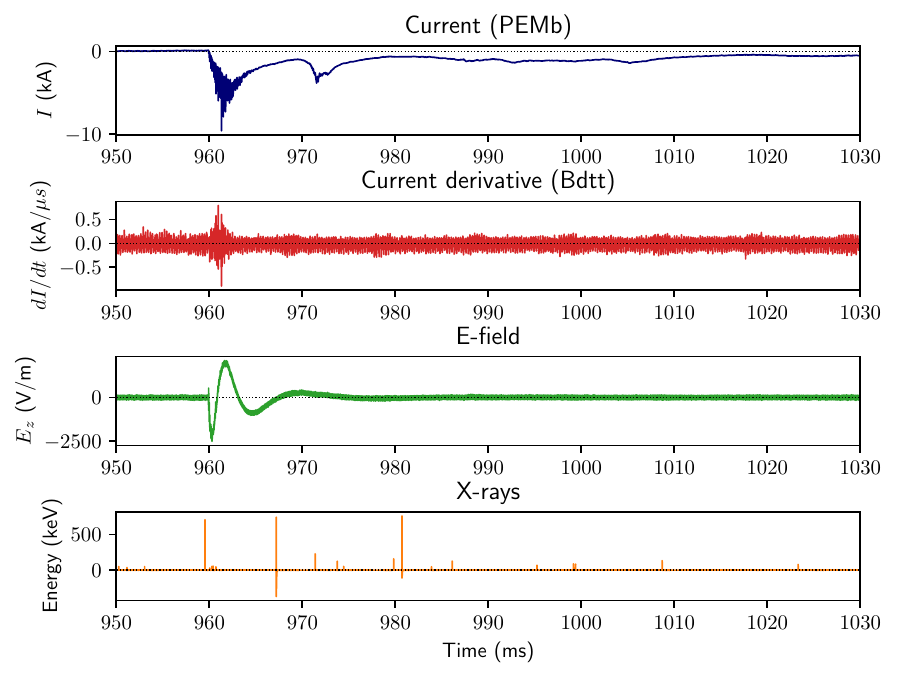}
        \subcaption{The entire duration of the flash. A 100 kHz low-pass filter has been applied to the current and $dI/dt$ waveforms to remove intermittent noise.}\label{fig:up3w}
    \end{subfigure}
    \hfill
    \begin{subfigure}[t]{0.49\textwidth}
        \centering
        \includegraphics[width=\textwidth]{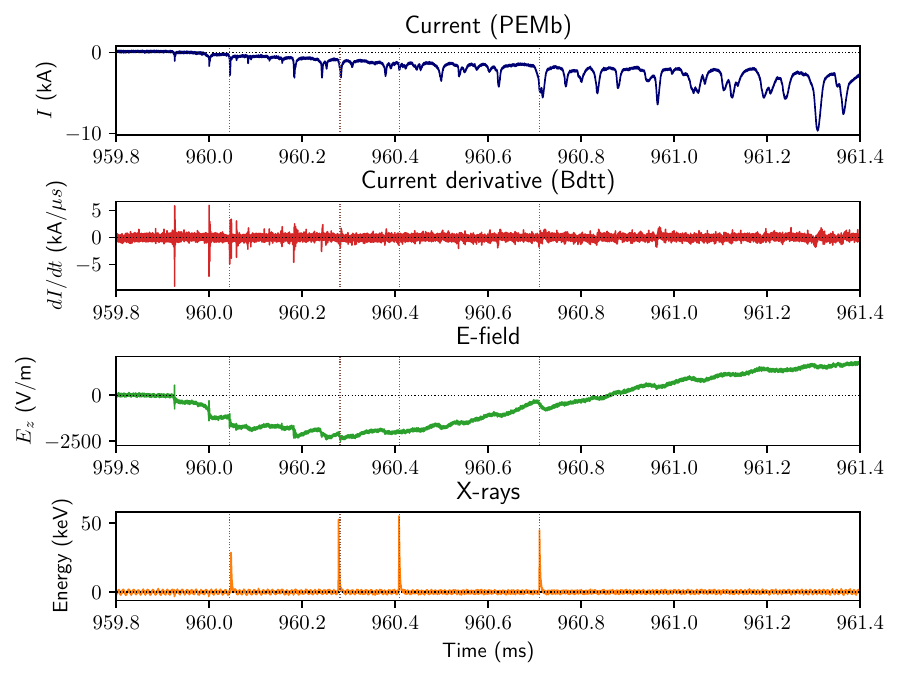}
        \subcaption{Zoom on the X-ray events during the upward stepping negative leader phase. The brown vertical dotted lines indicate the event times. See Table~\ref{tab:pfxr} for pulse data.}\label{fig:up3sl}
    \end{subfigure}
    \caption{Data associated with the Type 2 upward positive flash UP3, that occurred on July 30, 2021 at 18:00:10 UTC. ``PEMb'' and ``Bdtt'' specify the bottom Rogowski coil and top $\dot{B}$ sensor, respectively. $E_z$ is the measured vertical component of the electric field. The time is from the beginning of the recording ($\sim$1 second before the current peak).}
    \label{fig:UP3}
\end{figure}

%-------------------------------%

%An appendix contains supplementary information that is not an essential part of the text itself but which may be helpful in providing a more comprehensive understanding of the research problem or it is information that is too cumbersome to be included in the body of the paper.




\end{appendices}


\bibliography{upxr}% common bib file


\begin{thebibliography}{19}
\ifx \bisbn   \undefined \def \bisbn  #1{ISBN #1}\fi
\ifx \binits  \undefined \def \binits#1{#1}\fi
\ifx \bauthor  \undefined \def \bauthor#1{#1}\fi
\ifx \batitle  \undefined \def \batitle#1{#1}\fi
\ifx \bjtitle  \undefined \def \bjtitle#1{#1}\fi
\ifx \bvolume  \undefined \def \bvolume#1{\textbf{#1}}\fi
\ifx \byear  \undefined \def \byear#1{#1}\fi
\ifx \bissue  \undefined \def \bissue#1{#1}\fi
\ifx \bfpage  \undefined \def \bfpage#1{#1}\fi
\ifx \blpage  \undefined \def \blpage #1{#1}\fi
\ifx \burl  \undefined \def \burl#1{\textsf{#1}}\fi
\ifx \doiurl  \undefined \def \doiurl#1{\url{https://doi.org/#1}}\fi
\ifx \betal  \undefined \def \betal{\textit{et al.}}\fi
\ifx \binstitute  \undefined \def \binstitute#1{#1}\fi
\ifx \binstitutionaled  \undefined \def \binstitutionaled#1{#1}\fi
\ifx \bctitle  \undefined \def \bctitle#1{#1}\fi
\ifx \beditor  \undefined \def \beditor#1{#1}\fi
\ifx \bpublisher  \undefined \def \bpublisher#1{#1}\fi
\ifx \bbtitle  \undefined \def \bbtitle#1{#1}\fi
\ifx \bedition  \undefined \def \bedition#1{#1}\fi
\ifx \bseriesno  \undefined \def \bseriesno#1{#1}\fi
\ifx \blocation  \undefined \def \blocation#1{#1}\fi
\ifx \bsertitle  \undefined \def \bsertitle#1{#1}\fi
\ifx \bsnm \undefined \def \bsnm#1{#1}\fi
\ifx \bsuffix \undefined \def \bsuffix#1{#1}\fi
\ifx \bparticle \undefined \def \bparticle#1{#1}\fi
\ifx \barticle \undefined \def \barticle#1{#1}\fi
\bibcommenthead
\ifx \bconfdate \undefined \def \bconfdate #1{#1}\fi
\ifx \botherref \undefined \def \botherref #1{#1}\fi
\ifx \url \undefined \def \url#1{\textsf{#1}}\fi
\ifx \bchapter \undefined \def \bchapter#1{#1}\fi
\ifx \bbook \undefined \def \bbook#1{#1}\fi
\ifx \bcomment \undefined \def \bcomment#1{#1}\fi
\ifx \oauthor \undefined \def \oauthor#1{#1}\fi
\ifx \citeauthoryear \undefined \def \citeauthoryear#1{#1}\fi
\ifx \endbibitem  \undefined \def \endbibitem {}\fi
\ifx \bconflocation  \undefined \def \bconflocation#1{#1}\fi
\ifx \arxivurl  \undefined \def \arxivurl#1{\textsf{#1}}\fi
\csname PreBibitemsHook\endcsname

%%% 1
\bibitem{moore_energetic_2001}
\begin{barticle}
\bauthor{\bsnm{Moore}, \binits{C.B.}},
\bauthor{\bsnm{Eack}, \binits{K.B.}},
\bauthor{\bsnm{Aulich}, \binits{G.D.}},
\bauthor{\bsnm{Rison}, \binits{W.}}:
\batitle{Energetic radiation associated with lightning stepped-leaders}.
\bjtitle{Geophys. Res. Lett.}
\bvolume{28}(\bissue{11}),
\bfpage{2141}--\blpage{2144}
(\byear{2001}).
\doiurl{10.1029/2001GL013140}.
Accessed 2022-09-30
\end{barticle}
\endbibitem

%%% 2
\bibitem{gurevich_runaway_1992}
\begin{barticle}
\bauthor{\bsnm{Gurevich}, \binits{A.V.}},
\bauthor{\bsnm{Milikh}, \binits{G.M.}},
\bauthor{\bsnm{Roussel-Dupre}, \binits{R.}}:
\batitle{Runaway electron mechanism of air breakdown and preconditioning during
  a thunderstorm}.
\bjtitle{Physics Letters A}
\bvolume{165}(\bissue{5-6}),
\bfpage{463}--\blpage{468}
(\byear{1992}).
\doiurl{10.1016/0375-9601(92)90348-P}.
Accessed 2023-03-22
\end{barticle}
\endbibitem

%%% 3
\bibitem{dwyer_energetic_2003}
\begin{barticle}
\bauthor{\bsnm{Dwyer}, \binits{J.R.}},
\bauthor{\bsnm{Uman}, \binits{M.A.}},
\bauthor{\bsnm{Rassoul}, \binits{H.K.}},
\bauthor{\bsnm{Al-Dayeh}, \binits{M.}},
\bauthor{\bsnm{Caraway}, \binits{L.}},
\bauthor{\bsnm{Jerauld}, \binits{J.}},
\bauthor{\bsnm{Rakov}, \binits{V.A.}},
\bauthor{\bsnm{Jordan}, \binits{D.M.}},
\bauthor{\bsnm{Rambo}, \binits{K.J.}},
\bauthor{\bsnm{Corbin}, \binits{V.}},
\bauthor{\bsnm{Wright}, \binits{B.}}:
\batitle{Energetic {Radiation} {Produced} {During} {Rocket}-{Triggered}
  {Lightning}}.
\bjtitle{Science}
\bvolume{299}(\bissue{5607}),
\bfpage{694}--\blpage{697}
(\byear{2003}).
\doiurl{10.1126/science.1078940}.
Accessed 2022-10-06
\end{barticle}
\endbibitem

%%% 4
\bibitem{dwyer_x-ray_2005}
\begin{barticle}
\bauthor{\bsnm{Dwyer}, \binits{J.R.}},
\bauthor{\bsnm{Rassoul}, \binits{H.K.}},
\bauthor{\bsnm{Al-Dayeh}, \binits{M.}},
\bauthor{\bsnm{Caraway}, \binits{L.}},
\bauthor{\bsnm{Chrest}, \binits{A.}},
\bauthor{\bsnm{Wright}, \binits{B.}},
\bauthor{\bsnm{Kozak}, \binits{E.}},
\bauthor{\bsnm{Jerauld}, \binits{J.}},
\bauthor{\bsnm{Uman}, \binits{M.A.}},
\bauthor{\bsnm{Rakov}, \binits{V.A.}},
\bauthor{\bsnm{Jordan}, \binits{D.M.}},
\bauthor{\bsnm{Rambo}, \binits{K.J.}}:
\batitle{X-ray bursts associated with leader steps in cloud-to-ground
  lightning}.
\bjtitle{Geophys. Res. Lett.}
\bvolume{32}(\bissue{1}),
\bfpage{01803}
(\byear{2005}).
\doiurl{10.1029/2004GL021782}.
Accessed 2022-10-06
\end{barticle}
\endbibitem

%%% 5
\bibitem{saleh_properties_2009}
\begin{barticle}
\bauthor{\bsnm{Saleh}, \binits{Z.}},
\bauthor{\bsnm{Dwyer}, \binits{J.}},
\bauthor{\bsnm{Howard}, \binits{J.}},
\bauthor{\bsnm{Uman}, \binits{M.}},
\bauthor{\bsnm{Bakhtiari}, \binits{M.}},
\bauthor{\bsnm{Concha}, \binits{D.}},
\bauthor{\bsnm{Stapleton}, \binits{M.}},
\bauthor{\bsnm{Hill}, \binits{D.}},
\bauthor{\bsnm{Biagi}, \binits{C.}},
\bauthor{\bsnm{Rassoul}, \binits{H.}}:
\batitle{Properties of the {X}-ray emission from rocket-triggered lightning as
  measured by the {Thunderstorm} {Energetic} {Radiation} {Array} ({TERA})}.
\bjtitle{J. Geophys. Res.}
\bvolume{114}(\bissue{D17}),
\bfpage{17210}
(\byear{2009}).
\doiurl{10.1029/2008JD011618}.
Accessed 2023-05-22
\end{barticle}
\endbibitem

%%% 6
\bibitem{mallick_study_2012}
\begin{barticle}
\bauthor{\bsnm{Mallick}, \binits{S.}},
\bauthor{\bsnm{Rakov}, \binits{V.A.}},
\bauthor{\bsnm{Dwyer}, \binits{J.R.}}:
\batitle{A study of {X}-ray emissions from thunderstorms with emphasis on
  subsequent strokes in natural lightning: {X}-{RAY} {EMISSIONS} {FROM}
  {NATURAL} {LIGHTNING}}.
\bjtitle{J. Geophys. Res.}
\bvolume{117}(\bissue{D16}),
(\byear{2012}).
\doiurl{10.1029/2012JD017555}.
Accessed 2023-05-22
\end{barticle}
\endbibitem

%%% 7
\bibitem{yoshida_high_2008}
\begin{botherref}
\oauthor{\bsnm{Yoshida}, \binits{S.}},
\oauthor{\bsnm{Morimoto}, \binits{T.}},
\oauthor{\bsnm{Ushio}, \binits{T.}},
\oauthor{\bsnm{Kawasaki}, \binits{Z.-I.}},
\oauthor{\bsnm{Torii}, \binits{T.}},
\oauthor{\bsnm{Wang}, \binits{D.}},
\oauthor{\bsnm{Takagi}, \binits{N.}},
\oauthor{\bsnm{Watanabe}, \binits{T.}}:
High energy photon and electron bursts associated with upward lightning
  strokes: {HIGH} {ENERGY} {PHOTON} {AND} {ELECTRON} {BURSTS}.
Geophys. Res. Lett.
\textbf{35}(10)
(2008).
\doiurl{10.1029/2007GL032438}.
Accessed 2023-05-22
\end{botherref}
\endbibitem

%%% 8
\bibitem{montanya_registration_2014}
\begin{barticle}
\bauthor{\bsnm{Montanyà}, \binits{J.}},
\bauthor{\bsnm{Fabró}, \binits{F.}},
\bauthor{\bparticle{van~der} \bsnm{Velde}, \binits{O.}},
\bauthor{\bsnm{Romero}, \binits{D.}},
\bauthor{\bsnm{Solà}, \binits{G.}},
\bauthor{\bsnm{Hermoso}, \binits{J.R.}},
\bauthor{\bsnm{Soula}, \binits{S.}},
\bauthor{\bsnm{Williams}, \binits{E.R.}},
\bauthor{\bsnm{Pineda}, \binits{N.}}:
\batitle{Registration of {X}-rays at 2500 m altitude in association with
  lightning flashes and thunderstorms}.
\bjtitle{J. Geophys. Res. Atmos.}
\bvolume{119}(\bissue{3}),
\bfpage{1492}--\blpage{1503}
(\byear{2014}).
\doiurl{10.1002/2013JD021011}.
Accessed 2022-09-30
\end{barticle}
\endbibitem

%%% 9
\bibitem{hettiarachchi_x-ray_2018}
\begin{barticle}
\bauthor{\bsnm{Hettiarachchi}, \binits{P.}},
\bauthor{\bsnm{Cooray}, \binits{V.}},
\bauthor{\bsnm{Diendorfer}, \binits{G.}},
\bauthor{\bsnm{Pichler}, \binits{H.}},
\bauthor{\bsnm{Dwyer}, \binits{J.}},
\bauthor{\bsnm{Rahman}, \binits{M.}}:
\batitle{X-ray {Observations} at {Gaisberg} {Tower}}.
\bjtitle{Atmosphere}
\bvolume{9}(\bissue{1}),
\bfpage{20}
(\byear{2018}).
\doiurl{10.3390/atmos9010020}.
Accessed 2022-05-25
\end{barticle}
\endbibitem

%%% 10
\bibitem{sunjerga_santis_2021}
\begin{botherref}
\oauthor{\bsnm{Sunjerga}, \binits{A.}},
\oauthor{\bsnm{Mostajabi}, \binits{A.}},
\oauthor{\bsnm{Paolone}, \binits{M.}},
\oauthor{\bsnm{Rachidi}, \binits{F.}},
\oauthor{\bsnm{Romero}, \binits{C.}},
\oauthor{\bsnm{Hettiarachchi}, \binits{P.}},
\oauthor{\bsnm{Cooray}, \binits{V.}},
\oauthor{\bsnm{Azadifar}, \binits{M.}},
\oauthor{\bsnm{Rubinstein}, \binits{A.}},
\oauthor{\bsnm{Rubinstein}, \binits{M.}},
\oauthor{\bsnm{Pavanello}, \binits{D.}},
\oauthor{\bsnm{Smith}, \binits{D.}}:
Säntis {Lightning} {Research} {Facility} {Instrumentation}.
ICLP,
6
(2021)
\end{botherref}
\endbibitem

%%% 11
\bibitem{rachidi_santis_2022}
\begin{barticle}
\bauthor{\bsnm{Rachidi}, \binits{F.}},
\bauthor{\bsnm{Rubinstein}, \binits{M.}}:
\batitle{Säntis lightning research facility: a summary of the first ten years
  and future outlook}.
\bjtitle{Elektrotech. Inftech.}
\bvolume{139}(\bissue{3}),
\bfpage{379}--\blpage{394}
(\byear{2022}).
\doiurl{10.1007/s00502-022-01031-2}.
Accessed 2023-07-02
\end{barticle}
\endbibitem

%%% 12
\bibitem{sunjerga_initiation_2021}
\begin{botherref}
\oauthor{\bsnm{Sunjerga}, \binits{A.}},
\oauthor{\bsnm{Rubinstein}, \binits{M.}},
\oauthor{\bsnm{Rachidi}, \binits{F.}},
\oauthor{\bsnm{Cooray}, \binits{V.}}:
On the {Initiation} of {Upward} {Negative} {Lightning} by {Nearby} {Lightning}
  {Activity}: {An} {Analytical} {Approach}.
Geophys Res Atmos
\textbf{126}(5)
(2021).
\doiurl{10.1029/2020JD034043}.
Accessed 2022-05-16
\end{botherref}
\endbibitem

%%% 13
\bibitem{houard_laser-guided_2023}
\begin{barticle}
\bauthor{\bsnm{Houard}, \binits{A.}},
\bauthor{\bsnm{Walch}, \binits{P.}},
\bauthor{\bsnm{Produit}, \binits{T.}},
\bauthor{\bsnm{Moreno}, \binits{V.}},
\bauthor{\bsnm{Mahieu}, \binits{B.}},
\bauthor{\bsnm{Sunjerga}, \binits{A.}},
\bauthor{\bsnm{Herkommer}, \binits{C.}},
\bauthor{\bsnm{Mostajabi}, \binits{A.}},
\bauthor{\bsnm{Andral}, \binits{U.}},
\bauthor{\bsnm{André}, \binits{Y.-B.}},
\bauthor{\bsnm{Lozano}, \binits{M.}},
\bauthor{\bsnm{Bizet}, \binits{L.}},
\bauthor{\bsnm{Schroeder}, \binits{M.C.}},
\bauthor{\bsnm{Schimmel}, \binits{G.}},
\bauthor{\bsnm{Moret}, \binits{M.}},
\bauthor{\bsnm{Stanley}, \binits{M.}},
\bauthor{\bsnm{Rison}, \binits{W.A.}},
\bauthor{\bsnm{Maurice}, \binits{O.}},
\bauthor{\bsnm{Esmiller}, \binits{B.}},
\bauthor{\bsnm{Michel}, \binits{K.}},
\bauthor{\bsnm{Haas}, \binits{W.}},
\bauthor{\bsnm{Metzger}, \binits{T.}},
\bauthor{\bsnm{Rubinstein}, \binits{M.}},
\bauthor{\bsnm{Rachidi}, \binits{F.}},
\bauthor{\bsnm{Cooray}, \binits{V.}},
\bauthor{\bsnm{Mysyrowicz}, \binits{A.}},
\bauthor{\bsnm{Kasparian}, \binits{J.}},
\bauthor{\bsnm{Wolf}, \binits{J.-P.}}:
\batitle{Laser-guided lightning}.
\bjtitle{Nat. Photon.}
\bvolume{17}(\bissue{3}),
\bfpage{231}--\blpage{235}
(\byear{2023}).
\doiurl{10.1038/s41566-022-01139-z}.
Accessed 2023-08-16
\end{barticle}
\endbibitem

%%% 14
\bibitem{giri_relationship_2009}
\begin{botherref}
\oauthor{\bsnm{Giri}, \binits{D.V.}},
\oauthor{\bsnm{Prather}, \binits{W.D.}},
\oauthor{\bsnm{Baum}, \binits{C.E.}}:
The {Relationship} {Between} {NEMP} {Standards} and {Simulator} {Performance}
  {Specifications}.
Sensor and Simulation Notes
\textbf{Note 538}
(2009)
\end{botherref}
\endbibitem

%%% 15
\bibitem{romero_positive_2013}
\begin{barticle}
\bauthor{\bsnm{Romero}, \binits{C.}},
\bauthor{\bsnm{Rachidi}, \binits{F.}},
\bauthor{\bsnm{Rubinstein}, \binits{M.}},
\bauthor{\bsnm{Paolone}, \binits{M.}},
\bauthor{\bsnm{Rakov}, \binits{V.A.}},
\bauthor{\bsnm{Pavanello}, \binits{D.}}:
\batitle{Positive lightning flashes recorded on the {Säntis} tower from {May}
  2010 to {January} 2012: {POSITIVE} {LIGHTNING} {SÄNTIS} {TOWER}}.
\bjtitle{J. Geophys. Res. Atmos.}
\bvolume{118}(\bissue{23}),
\bfpage{12879}--\blpage{12892}
(\byear{2013}).
\doiurl{10.1002/2013JD020242}.
Accessed 2023-05-01
\end{barticle}
\endbibitem

%%% 16
\bibitem{rakov_lightning_2003}
\begin{bbook}
\bauthor{\bsnm{Rakov}, \binits{V.A.}},
\bauthor{\bsnm{Uman}, \binits{M.A.}}:
\bbtitle{Lightning: Physics and Effects}.
\bpublisher{Cambridge University Press},
\blocation{Cambridge, U.K. ; New York}
(\byear{2003})
\end{bbook}
\endbibitem

%%% 17
\bibitem{moss_monte_2006}
\begin{barticle}
\bauthor{\bsnm{Moss}, \binits{G.D.}},
\bauthor{\bsnm{Pasko}, \binits{V.P.}},
\bauthor{\bsnm{Liu}, \binits{N.}},
\bauthor{\bsnm{Veronis}, \binits{G.}}:
\batitle{Monte {Carlo} model for analysis of thermal runaway electrons in
  streamer tips in transient luminous events and streamer zones of lightning
  leaders}.
\bjtitle{J. Geophys. Res.}
\bvolume{111}(\bissue{A2}),
\bfpage{02307}
(\byear{2006}).
\doiurl{10.1029/2005JA011350}.
Accessed 2023-05-22
\end{barticle}
\endbibitem

%%% 18
\bibitem{dwyer_implications_2004}
\begin{barticle}
\bauthor{\bsnm{Dwyer}, \binits{J.R.}}:
\batitle{Implications of x-ray emission from lightning: {IMPLICATIONS} {OF}
  {X}-{RAYS} {FROM} {LIGHTNING}}.
\bjtitle{Geophys. Res. Lett.}
\bvolume{31}(\bissue{12}),
(\byear{2004}).
\doiurl{10.1029/2004GL019795}.
Accessed 2023-04-27
\end{barticle}
\endbibitem

%%% 19
\bibitem{cooray_electric_2009}
\begin{barticle}
\bauthor{\bsnm{Cooray}, \binits{V.}},
\bauthor{\bsnm{Becerra}, \binits{M.}},
\bauthor{\bsnm{Rakov}, \binits{V.}}:
\batitle{On the electric field at the tip of dart leaders in lightning
  flashes}.
\bjtitle{Journal of Atmospheric and Solar-Terrestrial Physics}
\bvolume{71}(\bissue{12}),
\bfpage{1397}--\blpage{1404}
(\byear{2009}).
\doiurl{10.1016/j.jastp.2009.06.002}.
Accessed 2023-09-26
\end{barticle}
\endbibitem

\end{thebibliography}


\end{document}